\documentstyle[12pt]{article}
\renewcommand\caption{}

\title{Chiral Symmetry Breaking in the Nambu--Jona-Lasinio Model\\
       in External Constant Electromagnetic Field}

\author{{\sl A.Yu. Babansky $^1$, E.V. Gorbar $^2$, and G.V. Shchepanyuk $^3$}\\
{\sl{
$^1$ Kiev State University, Ukraine}}\\
{\sl{
$^2$ Bogolyubov Institute for Theoretical Physics, Ukraine}}\\
{\sl{
$^3$ Institute of Mathematics, Ukraine}}
}

\date{}

\begin{document}

\maketitle

\vfill

\begin{abstract}
Dynamical chiral symmetry breaking (D$\chi$SB) is studied in the
Nambu--Jona-Lasinio model for an arbitrary combination of external
constant electric and magnetic fields.  In
$3+1$ dimensions it is shown that the critical coupling constant
increases with increasing of the value of the second invariant of
electromagentic field $\vec{E}\cdot\vec{B}$, i.e. the second invariant
inhibits $D\chi SB$.
The case of $2+1$ dimensions is simpler because there is only one
Lorentz invariant of electromagnetic field and any combination of
constant fields can be reduced to cases either purely magnetic or
purely electric field.
\end{abstract}

\vfill
\eject

\newpage
In works [1--3] it was first shown in the so called ladder
approximation that quantum electrodynamics (QED)
in the regime of strong
coupling has a new phase with dynamically broken chiral symmetry.
The new phase of QED possesses very interesting properties from the
theoretical viewpoint [4--6].  There were attempts to use them for
modelling electroweak symmetry breaking in technicolor-like models
[7].  However, at present the new phase of QED has little
relevance to experiment because we need a strong coupling constant $\alpha_c
\approx 1$ (recall that the physical value of electromagnetic
coupling constant is $\alpha_c=\frac{1}{137}\ll 1$).

However, as was suggested in [8, 9], the situation may
drastically change in the presence of strong external
electromagnetic fields where dynamical chiral symmetry breaking
$(D\chi SB)$ may occur at the regime of weak coupling.  A
breakthrough in this direction was made in [10, 11],
where, in the framework of the Nambu-Jona-Lasinio (NJL) model [12],
it was shown that $D\chi SB$ takes place in an external
constant magnetic field at any small attraction between fermions
both in $2+1$ and $3+1$ dimensions (note that the fact that
external magnetic field enhances $D\chi SB$ was first noted in the
NJL model in [13] (see also [14])).

As shown in [10, 11], in the infrared, the dynamics of fermions in
magnetic field in $2+1$ and $3+1$ dimensions resembles the dynamics
of fermions in $0+1$ and $1+1$ dimensions, respectively.
Therefore, we have an effective reduction of dimension of
space-time by 2 units and as a result the critical
value of the coupling coustant in external magnetic field is equal
to zero.  It was latter shown in [15] that the same effect takes
place in QED in external magnetic field.
Note that although the critical value of coupling constant is zero
extremely strong magnetic fields ($|\vec{B}| \ge 10^{13} G$) are
necessary for experimentally significant consequences because the
correction to the physical mass of electron is very tiny for weak
magnetic fields.

The case of constant electric field was considered in [13] where it
was shown that the value of the critical coupling constant is more
in this case than in the case without electric field.
In the present work we study D$\chi$SB in the NJL model in the case
of an arbitrary combination of constant electric and magnetic fields
in $3+1$ and $2+1$ dimensions.  We first consider the case of $3+1$
dimensions.
As well known, electromagnetic field has two Lorentz invariants
$f_1=\frac{1}{2}F_{\mu\nu} F^{\mu\nu} =\vec{B}^2-\vec{E}^2$ and
$f_2=\frac{1}{2}\varepsilon^{\mu\nu\alpha\beta}
F_{\mu\nu}F_{\alpha\beta}=\vec{E}\cdot\vec{B}$.  Since D$\chi$ SB was
already studied in cses of purely electric and magnetic constant
external fields when only the first invariant $f_1$ of electromagnetic
field is not equal to zero, in the present work we consider D$\chi$SB
in the case where $f_2 \ne 0$.

The Lagrangian of the NJL model [12] in an external
electromagnetic field reads
\begin{equation}
{\cal L} = \sum^N_{j=1} i\bar{\Psi}_j \gamma^\mu D_\mu
\Psi_j +\frac{G}{2} \sum^N_{j=1} \left[ (\bar{\Psi}_j\Psi_j)^2+
(\bar{\Psi}_ji\gamma_5\Psi_j)^2 \right],
\end{equation}
where $D_\mu$ is the covariant derivative $D_\mu=\partial_\mu+ie
A_\mu$ and $j=1,2,\dots, N$ flavor index.  Lagrangian (1) is
invariant with respect to the $U_L(N)\times U_R(N)$ chiral group.
By using auxiliary fields $\pi$ and $\sigma$, we can rewrite (1) in
the following form:
\begin{equation}
{\cal L}=\sum^N_{j=1} \left[i\bar{\Psi}_j \gamma^\mu D_\mu
 \Psi_j - \bar{\Psi}_j(\sigma_j+i\gamma_5\pi_j)\Psi_j
- \frac{1}{2G} \left(\sigma_j^2+\pi_j^2\right)\right]
\end{equation}

By taking integrals over fermion fields, we obtain the effective
action for $\pi$ and $\sigma$ fields
\begin{equation}
\Gamma(\sigma,\pi) =- i \sum^N_{j=1}\mbox{ Tr Ln}
\left[i\gamma^\mu D_\mu - (\sigma_{j}+i\gamma_5\pi_{j})\right] -
\frac{1}{2G}\int d^4x(\sigma^2_{j}+\pi^2_{j}).
\end{equation}

To obtain the effective potential for $\sigma$ and $\pi$ fields, it
suffices to consider the case of constant fields $\sigma =
const,$ $\pi = const$.  Since the effective action is invariant
with respect to the $U_L(N)\times U_R(N)$ chiral symmetry, the
effective potential depends on $\pi$ and $\sigma$ fields only
through the chirally invariant combination
$\rho^2=\sum^N_{j=1}(\sigma_j^2+\pi_j^2)$.  Therefore, in what
follows it is sufficient to set $\pi_k=0$ , $\sigma_k=0$ for
$k=2,\dots,N$ and consider the effective potential only for the
field $\sigma_1$ which we simply denote $\sigma$.  Thus,
\begin{equation}
\Gamma(\sigma) =- i\mbox {Tr Ln} \left[i\gamma^\mu D_\mu -
\sigma\right] -\frac{1}{2G}\int d^4x\sigma^2.
\end{equation}
By using the method of proper time [16, 17], we represent the first
term in (4) as follows:
\begin{equation}
-iTrLn (i D_{\mu}\gamma^{\mu} - \sigma) = -\frac{i}{2} Tr Ln (D^2 +
\sigma^2) =
\int \frac{i}{2s} tr \langle x|e^{-is(D^2+\sigma^2)}|x \rangle dsd^4x
\end{equation}

As well known [17], vacuum of QED is not stable in an external
electric field and the effective potential has an imaginary part
which defines the rate of birth of fermion-antifermion pairs from
vacuum per unit volume.  Since we study the problem of $D\chi SB$,
we can ignore this effect and consider only the real part of
effective potential which is equal to
\begin{equation}
V(\sigma)= \frac{\sigma^2}{2G}+\frac{N}{8\pi^2}v.p.\;
\int^\infty_{1/\Lambda^2} ds
\frac{1}{s} e^{-s\sigma^2}M\coth(Ms)L\cot(Ls),
\end{equation}
where $L^2 = e^2\frac{\sqrt{f_1^2 + 4f_2^2} - f_1}{2}$,
$M^2 = e^2\frac{\sqrt{f_1^2 + 4f_2^2} + f_1}{2}$, and
$f_1=\vec{B}^2 - \vec{E}^2$ and $f_2=\vec{E}\cdot\vec{B}$ are two
invariants of electromagnetic field.  In (6) we introduced
a cut-off $\frac{1}{\Lambda^2}$ and v.p. of the integral in s is present
because we consider only the real part of the effective potential
(recall that the imaginary part of the effective potential is given
by residues in poles of $\cot(Ls)$).  The gap equation $\delta
V/\delta\sigma|_{\sigma=m}=0$ has the form
\begin{equation}
\frac{1}{G}-\frac{N}{4\pi^2}v.p.\;\int^\infty_{1/\Lambda^2} ds
e^{-sm^2}M\coth(Ms)L\cot(Ls)=0.
\end{equation}

This gap equation was investigated in cases where only the first
invariant of
electromagnetic field is not equal to zero, i.e.
for cases of purely electric and magnetic external fields.  In this paper
we study how the presence of nonzero electric field
parallel to magnetic field  ($\vec{E}\cdot\vec{B} \ne 0$)
affects $D\chi SB$.  By using some inequalities, we first analytically
obtain an estimate from below for the critical coupling constant.  We add
and subtract 1/s to $M\coth(Ms)$ in the gap equation (7).
Then
\begin{eqnarray}
v.p.\;\int^\infty_{1/\Lambda^2} ds
e^{-sm^2}M\coth(Ms)L\cot(Ls) = \Lambda^2 - \frac{\pi}{2}L + \nonumber\\
v.p.\;\int^\infty_{1/\Lambda^2} ds
e^{-sm^2}(M\coth(Ms) - 1/s)L\cot(Ls),
\end{eqnarray}
where we used the result [13]
$v.p.\;\int^\infty_{1/\Lambda^2} ds
e^{-sm^2}L\frac{\cot(Ls)}{s} = \Lambda^2 - \frac{\pi}{2}L$.
Further, we represent the integral in (8) as a sum of two integrals
$\int_{1/\Lambda^2}^{\infty} = \int_{1/\Lambda^2}^{\frac{\pi}{2L}} +
\int_{\frac{\pi}{2L}}^{\infty}$ (note that $\frac{\pi}{2L}$
is the first zero of $\cot(Ls)$).
We now consider the integral from $\frac{\pi}{2L}$ to infinity.
Since coth\,x $\le$ 1/x + 1 for
$x > 0$, we have
\begin{equation}
v.p.\;\int_{\frac{\pi}{2L}}^\infty ds
e^{-sm^2}(M\coth(Ms) - 1/s)L\cot(Ls) \le
v.p.\;\int_{\frac{\pi}{2L}}^\infty ds
e^{-sm^2}ML\cot(Ls).
\end{equation}
Integrating by part, we obtain
\begin{eqnarray}
v.p.\;\int^\infty_{\frac{\pi}{2L}} ds
e^{-sm^2}ML\cot(Ls) = M\int^\infty_{\frac{\pi}{2L}} e^{-m^2s}
d(\mbox{ln}(2|\mbox{sin}Ls|))\nonumber\\
= -M\mbox{ln}(2|\sin\frac{\pi}{2}|) +
m^2M\int^\infty_{\frac{\pi}{2L}}
e^{-m^2s}\mbox{ln}(2|\mbox{sin}Ls|)ds.
\end{eqnarray}
By using the formula [18]
\begin{equation} \int^{\infty}_{0} e^{-qx}
\mbox{ln}(2|\mbox{sin}ax|) dx = -q \sum_{k=1}^{\infty}
\frac{1}{k(q^2 + 4k^2a^2)}, Re\,q > 0,
\end{equation}
and the fact that the integral
$M\int_{1/\Lambda^2}^{\frac{\pi}{2L}}\mbox{ln}(2|\mbox{sin}Ls|)ds$ is
finite, we conclude that the integral $m^2M\int^\infty_{\frac{\pi}{2L}}
e^{-m^2s}\mbox{ln}(2|\mbox{sin}Ls|)ds$ tends to zero on the
critical line (where $m^2 \to 0$).  Thus,
$v.p.\;\int^\infty_{\frac{\pi}{2L}} ds e^{-sm^2}ML\cot(Ls) =
-M\mbox{ln}2$.  Therefore, in view of (9), we have
\begin{equation}
v.p.\;\int_{\frac{\pi}{2L}}^\infty ds
e^{-sm^2}(M\coth(Ms) - 1/s)L\cot(Ls) \le -M\mbox{ln}2.
\end{equation}
It remains to estimate from below the integral
$\int_{1/\Lambda^2}^{\frac{\pi}{2L}} ds
(M\coth(Ms) - 1/s)L\cot(Ls)$.  We have
$\int^{\frac{\pi}{2L}}_{1/\Lambda^2} ds
(M\coth(Ms) - 1/s)L\cot(Ls) \le \int^{\frac{\pi}{2L}}_{1/\Lambda^2}
(M\cot(Ms) - 1/s) \frac{ds}{s}$ (because cot\,x $\le 1/x$ for x in the
interval from 0 to $\frac{\pi}{2}$).  If $M \gg L$, then we use
the estimate coth\,x $\le$ 1/x + 1 because $\coth(Ms)$ is
approximately 1 near the upper limit of integration.  Therefore,
in this case
\begin{equation}
\int^{\frac{\pi}{2L}}_{1/\Lambda^2} ds
(M\coth(Ms) - 1/s)L\cot(Ls) \le
M\mbox{ln}\frac{\pi\Lambda^2}{2L}.
\end{equation}
If $M \ll L$, then $\coth(Ms) \ll 1$ in the interval of integration.
By using the estimate coth\,x $\le
\frac{1}{x} + x/3$ ($\frac{1}{x}$ and x/3 are simply two first terms of
the Teylor expansion of cothx), we obtain
\begin{equation}
\int^{\frac{\pi}{2L}}_{1/\Lambda^2} ds
(M\coth(Ms) - 1/s)L\cot(Ls) \le
\frac{M^2}{3}(\frac{\pi}{2L} - 1/\Lambda^2).
\end{equation}
        We now analyse the obtained results.  We assume in what follows
that $|f_1| \gg |f_2|$.  In the magnetic-type
case ($f_1 > 0$), we have $L
\approx |e| (\frac{f_2^2}{f_1})^{1/2}$ and $M \approx |e|f_1^{1/2}$.
By using (8), (12), and (13), we obtain the following
estimate from below for the critical coupling constant in the
magnetic-type case:
\begin{equation}
g_{cr} \ge \frac{1}{1 -\frac{L\pi}{2\Lambda^2} +
\frac{M}{\Lambda^2}\mbox{ln}\frac{\Lambda^2}{2L} - \frac{M}{\Lambda^2}
\mbox{ln}2} \approx \frac{1}{1 +
|e|\frac{f_1^{1/2}}{\Lambda^2}\ln\frac{\Lambda^2f_1^{1/2}}{4|ef_2|}},
\end{equation}
where $g_{cr}$ is the dimensionless critical coupling constant
$g_{cr} = \frac{4\pi^2G\Lambda^2}{N}$.
It directly follows from (15) that
the presence of electric field parallel to magnetic field is very
important.  Indeed, if $f_2 \ne 0$, then $g_{cr}$
is no longer equal to zero (even if the magnetic field is very
strong $|\vec{B}| \sim \Lambda^2$) in contrast to the case of
purely magnetic field where $g_{cr} = 0$.  If $f_2$
increases, $g_{cr}$ is also increases.  If $f_2 \to 0$, then
$g_{cr} \to 0$, i.e. we recover the result obtained by Gusynin,
Miransky, and Shovkovy [10].  In the electric-type case
($f_1 < 0$) we have $L \approx |e||f_1|^{1/2}$ and
$M\approx |e|(\frac{f_2^2}{|f_1|})^{1/2}$.
By using (8), (12), and (14), we obtain
\begin{equation}
g_{cr} \ge \frac{1}{1 -\frac{L\pi}{2\Lambda^2} +
\frac{M^2}{3\Lambda^2}(\frac{\pi}{2L} - 1/\Lambda^2) -\frac{M}
{\Lambda^2}\mbox{ln}2}
\approx \frac{1}{1 -\frac{\pi|e||f_1|^{1/2}}{2\Lambda^2} -
\frac{|e|}{\Lambda^2}(\frac{f_2^2}{|f_1|})^{1/2}\mbox{ln}2}.
\end{equation}
It follows from (16) that in this case $g_{cr}$ for $f_2 \ne 0$ is
more than $g_{cr} = \frac{1}{1 -
\frac{\pi|e||f_1|^{1/2}}{2\Lambda^2}}$ [13] in the case
$f_2 = 0$.  As $f_2$ goes to zero, our estimate coincides with the
result obtained by Klevansky and Lemmer [13] in the electric-type
case.  Thus, we conclude from the obtained estimates that the
second invariant of electromagnetic field inhibits $D\chi SB$.
In magnetic-type case
it looks rather natural (indeed, if $f_2 \ne 0$, then it means
that $\vec{E} \ne 0$ and we know that
electric field inhibits $D\chi SB$).  However, in the
electric-type case it appears unlikely.  Indeed, let first $\vec{E}
\ne 0, \vec{B} = 0$ (therefore, $f_2 = 0$).
It is natural to assume that $g_{cr}$ should decrease in the case
$f_2 \ne 0$ because if $f_2 \ne 0$ it means that $\vec{B} \ne 0$
and we know that magnetic field assists $D\chi SB$.
We can understand the cause of growth of $g_{cr}$ with increasing
of $\vec{E}\cdot\vec{B}$ as follows.  Since we study the dependence
on the second invariant, we keep the first invariant $\vec{B}^2 -
\vec{E}^2$ unchanged.  Without loss of generality we can assume
that $\vec{E} \parallel \vec{B}$ (if not, one can perform an
appropriate Lorentz transformation).  If we increase $f_2$, then in
order to keep the first invariant unchanged we have to increase both
$\vec{B}$ and $\vec{E}$.  Therefore, there is a
competition between increasing of $\vec{B}$ and increasing of
$\vec{E}$.  It turned out that qualitatively increasing of
$\vec{E}$ is more significant for $g_{cr}$ than increasing of
$\vec{B}$ for any $f_1$, therefore, $g_{cr}$ always grows with
increasing of $f_2$.

We found a rather rough analytic estimate from below for the critical
coupling constant.  To obtain a more accurate dependence of the
critical coupling constant on $f_2$,
we numerically calculate the integral in (7).  A typical
dependence of $g_{cr}$ on
$f_2$ in the electric-type case is shown in Fig.1 (this figure corresponds
to $\frac{f_1}{\Lambda^4} = -10^{-4}$) and in the magnetic-type case
in Fig.2 (where $\frac{f_1}{\Lambda^4} = 10^{-4}$).

We see from these figures that the critical coupling
constant increases with increasing of $f_2$ for any value of $f_1$.
In the electric-type case
$g_{cr}$ is always more that 1.  In the magnetic-type case
$g_{cr}$ abruptly drops to zero as
$f_2$ tends to zero.  We also numerically calculated $g_{cr}$ in
the case where the first invariant is zero $f_1 = 0$
($|\vec{E}| = |\vec{B}|$) and obtained a
dependence which is similar to the electric-type case, i.e.
$g_{cr}$ increases with increasing of $f_2$.  Thus, the
numerical analysis of the gap equation confirms that the second
invariant of electromagnetic field inhibits $D\chi SB$.

We now consider the case of $2+1$ dimensions.  We use the
reducible 4-dimensional representation of the Dirac algebra for
fermion field in order that the model possess a chiral symmetry
(in fact, there is two chiral symmetries with $\gamma_5$ and
$\gamma_3$ matrices, for more details see [19]) and we do not
study parity breaking.  Thus, we have the following gap equation
for parity conserving mass:
\begin{equation}
m=2iG\mbox{tr}\big(S^{(m)}_A(x,x)\big),
\end{equation}
where $S^{(m)}_A$ is the fermion propagator in an external constant
electromagnetic field which has the following form in the
Fock--Schwinger proper time formalism:
\begin{equation}
S^{(m)}_A(x,x')=\big(i\partial_\mu\gamma^\mu  -eA_{\mu}\gamma^\mu -m \big)(-i)
\int^0_{-\infty}d\tau U_A(x,x';\tau),
\end{equation}
where $\epsilon \rightarrow 0$ and
\begin{eqnarray}
U_A(x,x';\tau)&=&<x|\frac{1}{(i\partial-eA)^2-\frac{e}{2}
\sigma^{\mu\nu}F_{\mu\nu}-m^2}|x'>
\nonumber\\
&=&\frac{e^{-i\frac{\pi}{4}}}{8\pi^{3/2}|\tau|^{3/2}}
\exp\Big[-ie\int^x_{x'}d\xi A(\xi)+
\nonumber\\
&~&
\frac{i}{4}(x-x')eF
\coth(eF\tau)(x-x')-
\nonumber\\
&~&\frac{1}{2}tr\big[\ln\big(\frac{\sinh eF\tau}{eF\tau}\big)\big]
+\frac{ie}{2}\sigma F\tau+im^2 \tau   \Big].
\end{eqnarray}
By taking trace over the Dirac indices in (13), we obtain
\begin{equation}
m=\frac{mG}{\pi^{3/2}}e^{-i\pi/4}\int^0_{-\infty}
\frac{d\tau}{|\tau|^{3/2}}(eX\tau)\cot(eX\tau),
\end{equation}
where
$X=\sqrt{\frac{1}{2}F_{\mu\nu}F^{\mu\nu}}=\sqrt{B^2-\vec{E}^2}$
(note that magnetic field is a pseudoscalar in $2+1$ dimensions,
not axial-vector).

As well known, in $2+1$ dimensions there is only one Lorentz
invariant $\frac{1}{2}F_{\mu\nu}F^{\mu\nu}$ of electromagnetic
field.  Indeed, we can see from (20) that the gap equation depends
on electromagnetic fields only through this Lorentz invariant.
Consequently, an arbitrary combination of constant fields can be
reduced to cases either purely magnetic or purely electric field.
The case of constant magnetic field in the Nambu--Jona--Lasinio
model in $2+1$ dimensions was considered in [10].  Therefore, we
study here only the case of constant electric field.

In the case of external electric field the gap equation takes the form
\begin{eqnarray}
m&=&\frac{mG}{\pi^{3/2}}e^{-i\pi/4} \int^\infty_{\frac{1}{\Lambda^2}}
\frac{ds}{s^{3/2}}e^{-sm^2}+
\nonumber\\
&~&\frac{mG}{\pi^{3/2}}e^{-i\pi/4} \int^0_{-\infty}
\frac{d\tau}{|\tau|^{3/2}}e^{i\tau(m^2-i\epsilon)}\big[(eE\tau)
\coth(eE\tau) - 1\big],
\end{eqnarray}
where we explicitly wrote down the term which corresponds to the
gap equation without external electric field and $E = |\vec{E}|$.
Further, by using the fact that\\ $\tau
\coth\tau-1=\tau\big(1+\frac{2}{e^{2\tau}-1}-\frac{1}{\tau}\big)$
and performing the change of variable $x=\frac{\tau}{2}$, for
$\mbox{Re}\Big[
e^{-i\pi/4}\int^0_{-\infty}\frac{d\tau}{|\tau|^{3/2}}
e^{i\tau(m^2-i\epsilon)}(eE\tau) \coth(eE\tau)\Big]$ we get
\begin{equation}
\mbox{Re}\Bigg[ I_E(\mu_E^2,\frac{1}{2})\Bigg]-\mbox{Im}
\Bigg[I_E(\mu_E^2,\frac{1}{2})\Bigg],
\end{equation}
where $I_E(\mu_E^2,\frac{1}{2})=\int^\infty_0 x^{a-1}
e^{i\frac{\mu_E^2}{2}x}\Big(\frac{1}{e^x-1}-\frac{1}{x}\Big)$,
$\mu_E^2=\frac{m_E^2}{eE}$ and we also used the equality
\begin{equation}
\frac{1}{2}\int^\infty_0\frac{dx}{x^{1/2}}\Big(\cos
\big(\frac{\mu_E^2}{2}x\big)-\sin\big(\frac{\mu_E^2}
{2}x\big)\Big)=0.
\end{equation}
$I_E(\mu_E^2,a)$ is an analytic function of a in the region
$0<\mbox{Re}(a)<2$.

For $1<\mbox{Re}(a)<2$, by representing $I_E(\mu_E^2)$ as a sum of
two integrals, we have
\begin{equation}
I_E(\mu_E^2, a)= -\Big(\frac{2}{\mu_E^2}\Big)^{a-1}
\Gamma(a-1)e^{i\frac{a-1}{2}\pi}+\Gamma(a)
\zeta\big(a,1-i\frac{\mu_E^2}{2}\big).
\end{equation}
Performing an analytic continuation of (24) to the region
$0<\mbox{Re}(a)<2$, we get
\begin{equation}
I_E\big(\mu_E^2,\frac{1}{2}\big)=\sqrt\pi\Big((1-i)\mu_E+
\zeta\big(\frac{1}{2},1-i\frac{\mu_E^2}{2}\big)\Big).
\end{equation}
Consequently, the gap equation takes the form
\begin{equation}
\frac{\pi^{3/2}}{2G\Lambda}=1+\frac{\sqrt\pi(eE)^{1/2}}{2\Lambda}\Bigg( \mbox{Re}
\Bigg[ \zeta \big(\frac12,1-i\frac{\mu_E^2}{2}\big)\Bigg]-\mbox{Im}
\Bigg[ \zeta \big(\frac12,1-i\frac{\mu_E^2}{2}\big)\Bigg]\Bigg) +
O\Big(\frac{m^2}{\Lambda^2}\Big).
\end{equation}
From (26), we obtain the following value for the critical coupling
constant:
\begin{eqnarray}
g_{cr}(E)=\frac{\pi^{3/2}}{2\Lambda}
\frac{1}{1-a^2\zeta(\frac{1}{2})(\frac{eE}{\Lambda^2})^{1/2}}
=\frac{g_{cr}(0)}{1-a^2(\frac{eE}{\Lambda^2})^{1/2}},
\end{eqnarray}
where $g_{cr} = \frac{2G\Lambda^2}{\pi^{3/2}}$ is the dimensionless
coupling constant in 2+1 dimension and
$a^2=-\frac{\sqrt\pi}{2}\zeta(\frac{1}{2})\approx 1.29$.

Thus, the presence  of external constant electric field increases the value of
critical coupling constant.
Note that the same is true in $3+1$ dimensions (see [13]).
It is not difficult to show that in the vicinity of critical point
\begin{equation}
m^2\approx C^2(E_{cr}-E)^{1/2},
\end{equation}
where $C^2=-2\frac{\zeta(1/2)}{\zeta(3/2)}\approx 1.12$.

Thus, this phase transition is a phase transition of
the second order.

Since we have only one Lorentz invariant of electromagnetic field
$X^2=\frac{1}{2}F_{\mu\nu}F_{\mu\nu}= {B}^2 - \vec{E}^2$ in $2+1$
dimensions, by using an appropriate frame, the general case of
non-zero constant electromagnetic field can be reduced to the cases
of purely electric or purely magnetic fields.

The authors are grateful to Prof. V.P. Gusynin for many fruitful
discussions and remarks, Prof. V.A. Miransky for valuable comments, and
Dr. I.A. Shovkovy for the help in drawing
figures.  The work of E.V.G. was supported in part
through grant INTAS-93-2058-EXT "East-West network in
constrained dynamical systems" and by the Foundation of
Fundamental Research of the Ministry of Science of the Ukraine through grant
No. 2.5.1/003.  The work of A.Yu.B. was supported
in part by the International Soros Science Education Program
(ISSEP) through grant No. GSU062075.

\newpage
\section*{Figure captions}
Figure 1: The dependence of the critical coupling constant on the
second invariant in the magnetic-type case.

\noindent
Figure 2: The dependence of the critical coupling constant on the
second invariant in the electric-type case.

\newpage

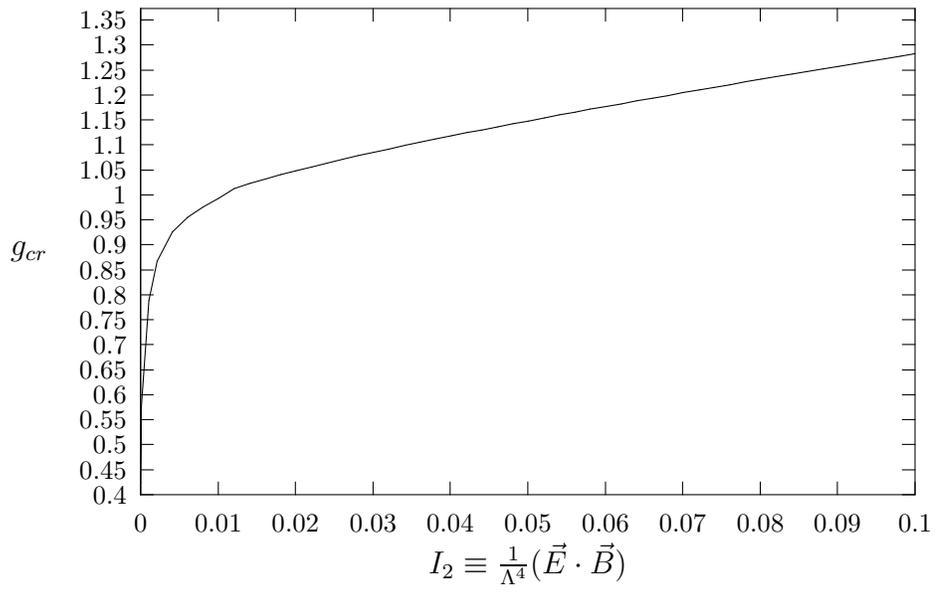
\begin{figure}
\setlength{\unitlength}{0.240900pt}
\ifx\plotpoint\undefined\newsavebox{\plotpoint}\fi
\sbox{\plotpoint}{\rule[-0.200pt]{0.400pt}{0.400pt}}%
\special{em:linewidth 0.4pt}%
\begin{picture}(1500,900)(0,0)
\font\gnuplot=cmr10 at 10pt
\gnuplot
\put(220,113){\special{em:moveto}}
\put(220,877){\special{em:lineto}}
\put(220,113){\special{em:moveto}}
\put(240,113){\special{em:lineto}}
\put(1436,113){\special{em:moveto}}
\put(1416,113){\special{em:lineto}}
\put(198,113){\makebox(0,0)[r]{0.4}}
\put(220,152){\special{em:moveto}}
\put(240,152){\special{em:lineto}}
\put(1436,152){\special{em:moveto}}
\put(1416,152){\special{em:lineto}}
\put(198,152){\makebox(0,0)[r]{0.45}}
\put(220,192){\special{em:moveto}}
\put(240,192){\special{em:lineto}}
\put(1436,192){\special{em:moveto}}
\put(1416,192){\special{em:lineto}}
\put(198,192){\makebox(0,0)[r]{0.5}}
\put(220,231){\special{em:moveto}}
\put(240,231){\special{em:lineto}}
\put(1436,231){\special{em:moveto}}
\put(1416,231){\special{em:lineto}}
\put(198,231){\makebox(0,0)[r]{0.55}}
\put(220,270){\special{em:moveto}}
\put(240,270){\special{em:lineto}}
\put(1436,270){\special{em:moveto}}
\put(1416,270){\special{em:lineto}}
\put(198,270){\makebox(0,0)[r]{0.6}}
\put(220,309){\special{em:moveto}}
\put(240,309){\special{em:lineto}}
\put(1436,309){\special{em:moveto}}
\put(1416,309){\special{em:lineto}}
\put(198,309){\makebox(0,0)[r]{0.65}}
\put(220,349){\special{em:moveto}}
\put(240,349){\special{em:lineto}}
\put(1436,349){\special{em:moveto}}
\put(1416,349){\special{em:lineto}}
\put(198,349){\makebox(0,0)[r]{0.7}}
\put(220,388){\special{em:moveto}}
\put(240,388){\special{em:lineto}}
\put(1436,388){\special{em:moveto}}
\put(1416,388){\special{em:lineto}}
\put(198,388){\makebox(0,0)[r]{0.75}}
\put(220,427){\special{em:moveto}}
\put(240,427){\special{em:lineto}}
\put(1436,427){\special{em:moveto}}
\put(1416,427){\special{em:lineto}}
\put(198,427){\makebox(0,0)[r]{0.8}}
\put(220,466){\special{em:moveto}}
\put(240,466){\special{em:lineto}}
\put(1436,466){\special{em:moveto}}
\put(1416,466){\special{em:lineto}}
\put(198,466){\makebox(0,0)[r]{0.85}}
\put(220,506){\special{em:moveto}}
\put(240,506){\special{em:lineto}}
\put(1436,506){\special{em:moveto}}
\put(1416,506){\special{em:lineto}}
\put(198,506){\makebox(0,0)[r]{0.9}}
\put(220,545){\special{em:moveto}}
\put(240,545){\special{em:lineto}}
\put(1436,545){\special{em:moveto}}
\put(1416,545){\special{em:lineto}}
\put(198,545){\makebox(0,0)[r]{0.95}}
\put(220,584){\special{em:moveto}}
\put(240,584){\special{em:lineto}}
\put(1436,584){\special{em:moveto}}
\put(1416,584){\special{em:lineto}}
\put(198,584){\makebox(0,0)[r]{1}}
\put(220,623){\special{em:moveto}}
\put(240,623){\special{em:lineto}}
\put(1436,623){\special{em:moveto}}
\put(1416,623){\special{em:lineto}}
\put(198,623){\makebox(0,0)[r]{1.05}}
\put(220,663){\special{em:moveto}}
\put(240,663){\special{em:lineto}}
\put(1436,663){\special{em:moveto}}
\put(1416,663){\special{em:lineto}}
\put(198,663){\makebox(0,0)[r]{1.1}}
\put(220,702){\special{em:moveto}}
\put(240,702){\special{em:lineto}}
\put(1436,702){\special{em:moveto}}
\put(1416,702){\special{em:lineto}}
\put(198,702){\makebox(0,0)[r]{1.15}}
\put(220,741){\special{em:moveto}}
\put(240,741){\special{em:lineto}}
\put(1436,741){\special{em:moveto}}
\put(1416,741){\special{em:lineto}}
\put(198,741){\makebox(0,0)[r]{1.2}}
\put(220,780){\special{em:moveto}}
\put(240,780){\special{em:lineto}}
\put(1436,780){\special{em:moveto}}
\put(1416,780){\special{em:lineto}}
\put(198,780){\makebox(0,0)[r]{1.25}}
\put(220,820){\special{em:moveto}}
\put(240,820){\special{em:lineto}}
\put(1436,820){\special{em:moveto}}
\put(1416,820){\special{em:lineto}}
\put(198,820){\makebox(0,0)[r]{1.3}}
\put(220,859){\special{em:moveto}}
\put(240,859){\special{em:lineto}}
\put(1436,859){\special{em:moveto}}
\put(1416,859){\special{em:lineto}}
\put(198,859){\makebox(0,0)[r]{1.35}}
\put(220,113){\special{em:moveto}}
\put(220,133){\special{em:lineto}}
\put(220,877){\special{em:moveto}}
\put(220,857){\special{em:lineto}}
\put(220,68){\makebox(0,0){0}}
\put(342,113){\special{em:moveto}}
\put(342,133){\special{em:lineto}}
\put(342,877){\special{em:moveto}}
\put(342,857){\special{em:lineto}}
\put(342,68){\makebox(0,0){0.01}}
\put(463,113){\special{em:moveto}}
\put(463,133){\special{em:lineto}}
\put(463,877){\special{em:moveto}}
\put(463,857){\special{em:lineto}}
\put(463,68){\makebox(0,0){0.02}}
\put(585,113){\special{em:moveto}}
\put(585,133){\special{em:lineto}}
\put(585,877){\special{em:moveto}}
\put(585,857){\special{em:lineto}}
\put(585,68){\makebox(0,0){0.03}}
\put(706,113){\special{em:moveto}}
\put(706,133){\special{em:lineto}}
\put(706,877){\special{em:moveto}}
\put(706,857){\special{em:lineto}}
\put(706,68){\makebox(0,0){0.04}}
\put(828,113){\special{em:moveto}}
\put(828,133){\special{em:lineto}}
\put(828,877){\special{em:moveto}}
\put(828,857){\special{em:lineto}}
\put(828,68){\makebox(0,0){0.05}}
\put(950,113){\special{em:moveto}}
\put(950,133){\special{em:lineto}}
\put(950,877){\special{em:moveto}}
\put(950,857){\special{em:lineto}}
\put(950,68){\makebox(0,0){0.06}}
\put(1071,113){\special{em:moveto}}
\put(1071,133){\special{em:lineto}}
\put(1071,877){\special{em:moveto}}
\put(1071,857){\special{em:lineto}}
\put(1071,68){\makebox(0,0){0.07}}
\put(1193,113){\special{em:moveto}}
\put(1193,133){\special{em:lineto}}
\put(1193,877){\special{em:moveto}}
\put(1193,857){\special{em:lineto}}
\put(1193,68){\makebox(0,0){0.08}}
\put(1314,113){\special{em:moveto}}
\put(1314,133){\special{em:lineto}}
\put(1314,877){\special{em:moveto}}
\put(1314,857){\special{em:lineto}}
\put(1314,68){\makebox(0,0){0.09}}
\put(1436,113){\special{em:moveto}}
\put(1436,133){\special{em:lineto}}
\put(1436,877){\special{em:moveto}}
\put(1436,857){\special{em:lineto}}
\put(1436,68){\makebox(0,0){0.1}}
\put(220,113){\special{em:moveto}}
\put(1436,113){\special{em:lineto}}
\put(1436,877){\special{em:lineto}}
\put(220,877){\special{em:lineto}}
\put(220,113){\special{em:lineto}}
\put(45,495){\makebox(0,0){$g_{cr}$}}
\put(828,5){\makebox(0,0){$I_2\equiv\frac{1}{\Lambda^4}({\vec{E}}\cdot{\vec{B}})$}}
\put(220,131){\special{em:moveto}}
\put(221,193){\special{em:lineto}}
\put(221,248){\special{em:lineto}}
\put(233,418){\special{em:lineto}}
\put(246,480){\special{em:lineto}}
\put(270,526){\special{em:lineto}}
\put(294,549){\special{em:lineto}}
\put(318,565){\special{em:lineto}}
\put(343,579){\special{em:lineto}}
\put(367,594){\special{em:lineto}}
\put(391,602){\special{em:lineto}}
\put(416,609){\special{em:lineto}}
\put(440,616){\special{em:lineto}}
\put(464,622){\special{em:lineto}}
\put(489,628){\special{em:lineto}}
\put(513,634){\special{em:lineto}}
\put(537,640){\special{em:lineto}}
\put(562,646){\special{em:lineto}}
\put(586,651){\special{em:lineto}}
\put(610,656){\special{em:lineto}}
\put(635,662){\special{em:lineto}}
\put(659,667){\special{em:lineto}}
\put(683,672){\special{em:lineto}}
\put(708,677){\special{em:lineto}}
\put(732,682){\special{em:lineto}}
\put(756,686){\special{em:lineto}}
\put(781,691){\special{em:lineto}}
\put(805,696){\special{em:lineto}}
\put(829,700){\special{em:lineto}}
\put(854,705){\special{em:lineto}}
\put(878,710){\special{em:lineto}}
\put(902,714){\special{em:lineto}}
\put(926,719){\special{em:lineto}}
\put(951,723){\special{em:lineto}}
\put(975,727){\special{em:lineto}}
\put(999,732){\special{em:lineto}}
\put(1024,736){\special{em:lineto}}
\put(1048,740){\special{em:lineto}}
\put(1072,745){\special{em:lineto}}
\put(1097,749){\special{em:lineto}}
\put(1121,753){\special{em:lineto}}
\put(1145,757){\special{em:lineto}}
\put(1170,762){\special{em:lineto}}
\put(1194,766){\special{em:lineto}}
\put(1218,770){\special{em:lineto}}
\put(1243,774){\special{em:lineto}}
\put(1267,778){\special{em:lineto}}
\put(1291,782){\special{em:lineto}}
\put(1316,786){\special{em:lineto}}
\put(1340,790){\special{em:lineto}}
\put(1364,794){\special{em:lineto}}
\put(1389,798){\special{em:lineto}}
\put(1413,802){\special{em:lineto}}
\put(1436,806){\special{em:lineto}}
\end{picture}
\centerline{\caption{Fig.1}}
\end{figure}

\begin{figure}
\setlength{\unitlength}{0.240900pt}
\ifx\plotpoint\undefined\newsavebox{\plotpoint}\fi
\sbox{\plotpoint}{\rule[-0.200pt]{0.400pt}{0.400pt}}%
\special{em:linewidth 0.4pt}%
\begin{picture}(1500,900)(0,0)
\font\gnuplot=cmr10 at 10pt
\gnuplot
\put(220,113){\special{em:moveto}}
\put(220,877){\special{em:lineto}}
\put(220,113){\special{em:moveto}}
\put(240,113){\special{em:lineto}}
\put(1436,113){\special{em:moveto}}
\put(1416,113){\special{em:lineto}}
\put(198,113){\makebox(0,0)[r]{0.85}}
\put(220,186){\special{em:moveto}}
\put(240,186){\special{em:lineto}}
\put(1436,186){\special{em:moveto}}
\put(1416,186){\special{em:lineto}}
\put(198,186){\makebox(0,0)[r]{0.9}}
\put(220,259){\special{em:moveto}}
\put(240,259){\special{em:lineto}}
\put(1436,259){\special{em:moveto}}
\put(1416,259){\special{em:lineto}}
\put(198,259){\makebox(0,0)[r]{0.95}}
\put(220,332){\special{em:moveto}}
\put(240,332){\special{em:lineto}}
\put(1436,332){\special{em:moveto}}
\put(1416,332){\special{em:lineto}}
\put(198,332){\makebox(0,0)[r]{1}}
\put(220,405){\special{em:moveto}}
\put(240,405){\special{em:lineto}}
\put(1436,405){\special{em:moveto}}
\put(1416,405){\special{em:lineto}}
\put(198,405){\makebox(0,0)[r]{1.05}}
\put(220,478){\special{em:moveto}}
\put(240,478){\special{em:lineto}}
\put(1436,478){\special{em:moveto}}
\put(1416,478){\special{em:lineto}}
\put(198,478){\makebox(0,0)[r]{1.1}}
\put(220,551){\special{em:moveto}}
\put(240,551){\special{em:lineto}}
\put(1436,551){\special{em:moveto}}
\put(1416,551){\special{em:lineto}}
\put(198,551){\makebox(0,0)[r]{1.15}}
\put(220,624){\special{em:moveto}}
\put(240,624){\special{em:lineto}}
\put(1436,624){\special{em:moveto}}
\put(1416,624){\special{em:lineto}}
\put(198,624){\makebox(0,0)[r]{1.2}}
\put(220,697){\special{em:moveto}}
\put(240,697){\special{em:lineto}}
\put(1436,697){\special{em:moveto}}
\put(1416,697){\special{em:lineto}}
\put(198,697){\makebox(0,0)[r]{1.25}}
\put(220,770){\special{em:moveto}}
\put(240,770){\special{em:lineto}}
\put(1436,770){\special{em:moveto}}
\put(1416,770){\special{em:lineto}}
\put(198,770){\makebox(0,0)[r]{1.3}}
\put(220,843){\special{em:moveto}}
\put(240,843){\special{em:lineto}}
\put(1436,843){\special{em:moveto}}
\put(1416,843){\special{em:lineto}}
\put(198,843){\makebox(0,0)[r]{1.35}}
\put(220,113){\special{em:moveto}}
\put(220,133){\special{em:lineto}}
\put(220,877){\special{em:moveto}}
\put(220,857){\special{em:lineto}}
\put(220,68){\makebox(0,0){0}}
\put(342,113){\special{em:moveto}}
\put(342,133){\special{em:lineto}}
\put(342,877){\special{em:moveto}}
\put(342,857){\special{em:lineto}}
\put(342,68){\makebox(0,0){0.01}}
\put(463,113){\special{em:moveto}}
\put(463,133){\special{em:lineto}}
\put(463,877){\special{em:moveto}}
\put(463,857){\special{em:lineto}}
\put(463,68){\makebox(0,0){0.02}}
\put(585,113){\special{em:moveto}}
\put(585,133){\special{em:lineto}}
\put(585,877){\special{em:moveto}}
\put(585,857){\special{em:lineto}}
\put(585,68){\makebox(0,0){0.03}}
\put(706,113){\special{em:moveto}}
\put(706,133){\special{em:lineto}}
\put(706,877){\special{em:moveto}}
\put(706,857){\special{em:lineto}}
\put(706,68){\makebox(0,0){0.04}}
\put(828,113){\special{em:moveto}}
\put(828,133){\special{em:lineto}}
\put(828,877){\special{em:moveto}}
\put(828,857){\special{em:lineto}}
\put(828,68){\makebox(0,0){0.05}}
\put(950,113){\special{em:moveto}}
\put(950,133){\special{em:lineto}}
\put(950,877){\special{em:moveto}}
\put(950,857){\special{em:lineto}}
\put(950,68){\makebox(0,0){0.06}}
\put(1071,113){\special{em:moveto}}
\put(1071,133){\special{em:lineto}}
\put(1071,877){\special{em:moveto}}
\put(1071,857){\special{em:lineto}}
\put(1071,68){\makebox(0,0){0.07}}
\put(1193,113){\special{em:moveto}}
\put(1193,133){\special{em:lineto}}
\put(1193,877){\special{em:moveto}}
\put(1193,857){\special{em:lineto}}
\put(1193,68){\makebox(0,0){0.08}}
\put(1314,113){\special{em:moveto}}
\put(1314,133){\special{em:lineto}}
\put(1314,877){\special{em:moveto}}
\put(1314,857){\special{em:lineto}}
\put(1314,68){\makebox(0,0){0.09}}
\put(1436,113){\special{em:moveto}}
\put(1436,133){\special{em:lineto}}
\put(1436,877){\special{em:moveto}}
\put(1436,857){\special{em:lineto}}
\put(1436,68){\makebox(0,0){0.1}}
\put(220,113){\special{em:moveto}}
\put(1436,113){\special{em:lineto}}
\put(1436,877){\special{em:lineto}}
\put(220,877){\special{em:lineto}}
\put(220,113){\special{em:lineto}}
\put(45,495){\makebox(0,0){$g_{cr}$}}
\put(828,5){\makebox(0,0){$I_2\equiv\frac{1}{\Lambda^4}({\vec{E}}\cdot{\vec{B}})$}}
\put(221,353){\special{em:moveto}}
\put(246,394){\special{em:lineto}}
\put(270,419){\special{em:lineto}}
\put(294,439){\special{em:lineto}}
\put(318,456){\special{em:lineto}}
\put(343,471){\special{em:lineto}}
\put(367,486){\special{em:lineto}}
\put(391,499){\special{em:lineto}}
\put(416,512){\special{em:lineto}}
\put(440,524){\special{em:lineto}}
\put(464,536){\special{em:lineto}}
\put(489,547){\special{em:lineto}}
\put(513,558){\special{em:lineto}}
\put(537,569){\special{em:lineto}}
\put(562,579){\special{em:lineto}}
\put(586,589){\special{em:lineto}}
\put(610,599){\special{em:lineto}}
\put(635,609){\special{em:lineto}}
\put(659,618){\special{em:lineto}}
\put(683,628){\special{em:lineto}}
\put(708,637){\special{em:lineto}}
\put(732,646){\special{em:lineto}}
\put(756,655){\special{em:lineto}}
\put(781,664){\special{em:lineto}}
\put(805,672){\special{em:lineto}}
\put(829,681){\special{em:lineto}}
\put(854,690){\special{em:lineto}}
\put(878,698){\special{em:lineto}}
\put(902,706){\special{em:lineto}}
\put(926,715){\special{em:lineto}}
\put(951,723){\special{em:lineto}}
\put(975,731){\special{em:lineto}}
\put(999,739){\special{em:lineto}}
\put(1024,747){\special{em:lineto}}
\put(1048,755){\special{em:lineto}}
\put(1072,763){\special{em:lineto}}
\put(1097,771){\special{em:lineto}}
\put(1121,779){\special{em:lineto}}
\put(1145,787){\special{em:lineto}}
\put(1170,795){\special{em:lineto}}
\put(1194,802){\special{em:lineto}}
\put(1218,810){\special{em:lineto}}
\put(1243,818){\special{em:lineto}}
\put(1267,825){\special{em:lineto}}
\put(1291,833){\special{em:lineto}}
\put(1316,840){\special{em:lineto}}
\put(1340,848){\special{em:lineto}}
\put(1364,856){\special{em:lineto}}
\put(1389,863){\special{em:lineto}}
\put(1413,870){\special{em:lineto}}
\put(1434,877){\special{em:lineto}}
\end{picture}
\centerline{\caption{Fig.2}}
\end{figure}

\end{document}